# Magnetic proximity-induced non-relativistic valley polarization

Weifeng Xie[1], Yongqi Chen[1], Yiboyang Sun[1], Xiong Xu[2], Yunliang Yue[3], Hui Wang[2*]

1 School of Microelectronics and Physics, Hunan University of Technology and Business, Changsha 410205, China
2 School of Physics, Hunan Key Laboratory of Super Microstructure and Ultrafast Process, Hunan Key Laboratory of Nanophotonics and Devices, State Key Laboratory of Powder Metallurgy, Central South University, Changsha 410083, China
3 School of Information and Artificial Intelligence, Yangzhou University, Yangzhou 225127, China

## Abstract

The magnetic proximity effect in van der Waals (vdW) heterostructures exerts a significant impact on the electronic and magnetic properties of adjacent materials. Here, we propose vdW heterostructures composed of monolayers ferromagnets (FM) and altermagnets (AM), in which the magnetic proximity effect from the FM induces pronounced non-relativistic valley polarization in the AM. Furthermore, by flipping the magnetization of the FM and the direction of Néel vector of the AM, four independent valley-polarized states can be realized in the FM/AM vdW heterostructures, exhibiting strong magneto-valley coupling. The phenomenon is qualitatively elucidated by a tight-binding model and demonstrated to be generalizable. These findings suggest that FM/AM vdW heterostructures hold potential application in the field of valleytronics-based information storage.

## Introduction

Utilizing the valley degree of freedom for information storage benefits from the inherently weak intervalley scattering in certain two-dimensional (2D) materials, which extends the lifetime of valley-polarized carriers and their coherence time, and thereby provides potential complementary advantages over charge and spin degrees of freedom for information storage[1]. Considerable valley polarization is a prerequisite for the practical application of the valley degree of freedom[1,2]. It has been demonstrated that spin-orbital coupling (SOC) will induce valley splitting at the K and K′ points in hexagonal graphene[3] and 2D non-magnetic transition metal dichalcogenides (TMD)[4–6]. However, intrinsically, due to the existence of time-reversal symmetry, the energy levels at different valleys remain degenerate. Therefore, external magnetic fields[7–9], circularly polarized light excitation[5,10,11], magnetic atom doping[12,13], and magnetic proximity effects from ferromagnetic substrates[14–16] are prerequisites for generating valley polarization in such systems. Subsequently, it is unveiled that in ferromagnetic TMD, intrinsic magnetization is equivalent to the effect of an external magnetic field, forming an intrinsic valley polarization correlated with the direction of magnetization, such materials are termed ferrovalley materials[17]. It should be noted that the formation of valley polarization in ferromagnetic ferrovalley materials still requires consideration of the SOC. Until recently, a class of collinear magnetic materials—altermagnets (AM)—has attracted widespread attention[18–21]. Based on the latest spin space group theory, studies have revealed that monolayer AM with a tetragonal lattice ($[C_2||C_{4z}]$) exhibit valley splitting at the X and Y points without considering SOC owing to their unique rotational symmetry and diagonal mirror symmetry[18,19,21]. The symmetry conditions enable significant non-relativistic valley polarization at the X and Y points solely through the application of uniaxial strain, a phenomenon referred to as the piezovalley effect[22–24]. Since valley polarization in AM does not rely on SOC, remarkable valley polarization can be induced by experimentally accessible uniaxial strain in compounds composed of light elements. Consequently, monolayer AM hold considerable potential for applications in the field of valleytronics.

In $MoS_2$-type non-magnetic TMD, valley splitting arises at different valleys under SOC, whereas AM intrinsically exhibit non-relativistic valley splitting, the two manifest identically in their band structures despite the different microscopic origins of the splitting[18]. In the former, valley polarization can be generated via external magnetic fields, magnetic atom doping and magnetic proximity effects[25,26]. Similarly, in principle, the latter can also regulate the bands of opposite spins through analogous methods, enabling non-relativistic valley polarization at valley points near the Fermi level. Compared with external magnetic fields and magnetic atom doping, constructing van der Waals (vdW) heterostructures composed of ferromagnets (FM) and AM to induce valley polarization in AM via the magnetic proximity effect of FM presents greater potential for exploration both theoretically and experimentally. This is primarily attributed to the diversity and flexibility in constructing vdW heterostructures, as well as the inheritance and integration of functionalities[27]. As is well known, semiconductor materials are more conducive to device integration. To date, most theoretically identified monolayer AM with a tetragonal lattice exhibit a characteristic of semiconductor[20,28,29]. Therefore, it is necessary to identify tetragonal ferromagnetic semiconductors that are lattice-matched with known monolayer AM, so as to reduce band hybridization at valley points near the Fermi level. However, investigations reveal that tetragonal ferromagnetic semiconductors are exceedingly rare, leaving FM/AM vdW heterostructures largely unexplored, and relevant phenomena of valley polarization remain scarcely investigated.

In this work, we explore vdW heterostructures formed by tetragonal ferromagnetic semiconductors CrP and CrAs with various monolayer AM, where the magnetic proximity effect from the FM is utilized to induce non-relativistic valley polarization in the AM. Our results from first-principles calculations and analysis from tight binding model reveal that independently tuning the magnetization of the FM and the Néel vector of the AM enables the AM to achieve four distinct valley-polarized states, demonstrating potential applications of FM/AM vdW heterostructures in valleytronics-based magnetic storage.

## Computational Methods

All calculations in this work are performed using the Vienna ab initio Simulation

Package (VASP) based on density functional theory (DFT)[30]. The projector-augmented wave (PAW) method[31] and the Perdew-Burke-Ernzerhof (PBE) functional within the generalized gradient approximation (GGA)[32] are adopted to describe the electron-ion interaction and exchange-correlation effects, respectively. A uniform plane-wave cutoff energy of 560 eV is employed. A vacuum layer larger than 15 Å along the direction perpendicular to the 2D plane is applied to avoid interlayer interactions affecting the calculation results. The energy convergence criteria for both structural optimization and self-consistent calculations are uniformly set to $1 \times 10^{-6}$ eV, and the convergence threshold for the force on each atom is set to 0.01 eV/Å. For the *k*-point sampling, a $12 \times 12 \times 1$ Γ-centered Monkhorst–Pack grid is used. To further account for the strong correlation effects of the *d* electrons of V and Cr, the PBE + *U* method[33] is uniformly adopted, with $U_{\text{eff}}$ set to 4 eV for V[23] and 3 eV for Cr[34–36]. Meanwhile, the DFT-D3 method[37] is employed to correct the interlayer vdW interactions in the heterostructures.

## Results and Discussion

Fig. 1 (a) and (b) show the lattice schematics of the 2D tetragonal AM, where the Néel vector is defined along the $+c$ and $-c$ axis, respectively. Because the AM share characteristics with antiferromagnets, reversing the Néel vector flips the local magnetic moments on the different sublattices. With the direction of spins taken along the $+c$ ($+z$) direction, here, we define a spin-resolved tight-binding model for the pure 2D AM as follows[38–40],

$$\mathcal{H}_s(\boldsymbol{k}) = -4t_1\cos\frac{k_x}{2}\cos\frac{k_y}{2}\tau_x - 2t_2\left(\cos k_x + \cos k_y\right)\tau_0$$
$$-2t_d\left(\cos k_x - \cos k_y\right)\tau_z + sJN\tau_z, \quad (1)$$

where $t_1$ indicates the nearest-neighbor hopping, $t_2$ and $t_d$ represent the isotropic and anisotropic next-nearest-neighbor hopping, respectively, that is $t_{2,d} = (t_{2a} \pm t_{2b})/2$, as illustrated in Fig. 1 (a) and (b), $\tau_i$ $(i = x, y, z)$ acts on the sublattice degrees of freedom ($\tau_0$ is unit matrix), and $J$ is the effective exchange coupling, the Néel vector $\boldsymbol{N}$ characterizes the collinear antiferromagnetic order, $s = +1$ and -1 represent spin-up (↑) and spin-down (↓) electrons, respectively. From Eq. (1), the Hamiltonian at the X and Y valleys can be written as

$$\mathcal{H}_s(\mathrm{X}) = \mathcal{H}_s(\pi, 0) = 4t_d\tau_z + sJN\tau_z, \quad (2)$$
$$\mathcal{H}_s(\mathrm{Y}) = \mathcal{H}_s(0, \pi) = -4t_d\tau_z + sJN\tau_z. \quad (3)$$

Then, it can be derived from Eq. (2) and (3) that

$$E_\uparrow(\mathrm{X}) - E_\uparrow(\mathrm{Y}) = 8t_d, \quad (4)$$

which corresponds to the schematic band structure of the AM shown in Fig. 1 (c). From Fig. 1 (c), it can be found that the intrinsic non-relativistic spin splitting in the AM exhibits spin-valley locking characteristics[18], resembling the band structure of non-magnetic TMD under SOC.

After the monolayers FM and AM form a vdW heterostructure, wave-function overlap across the vdW interface between the FM and AM gives rise to an interfacial exchange coupling $J_p$. The FM possesses a net magnetization $\mathbf{M}$, through the magnetic proximity effect, it induces a uniform effective Zeeman field $\boldsymbol{B}_{eff}$ in the spin space of electrons of the AM, defined as $\boldsymbol{B}_{eff} \propto J_p \mathbf{M}$. Therefore, the total Hamiltonian of the AM can be written as

$$\mathcal{H}_{tot}(\boldsymbol{k}) = \mathcal{H}_s(\boldsymbol{k}) - sB_{eff}\tau_0. \quad (5)$$

At the X (Y) valley of the AM, the spin is denoted as $s_{\mathrm{X}}$ ($s_{\mathrm{Y}}$), the spin at the Y valley satisfies $s_{\mathrm{Y}} = -s_{\mathrm{X}}$. According to Eq. (2), (3) and (5), it follows that the valley polarization ($\Delta E_c$) of the lowermost conduction band (LCB) of the AM induced by the magnetic proximity effect can be deduced as

$$\Delta E_c = E_{LCB}^{\uparrow}(\mathrm{X}) - E_{LCB}^{\downarrow}(\mathrm{Y}) = -2s_{\mathrm{X}}B_{eff}, \quad (6)$$

corresponding to Fig. 1 (d). It is concluded that when monolayer AM are assembled into vdW heterostructures with FM, the magnetic proximity effect from the FM shifts the opposite-spin bands of the LCB at the valleys in AM in opposite directions, thereby lifting the energy degeneracy at different valleys and giving rise to a non-relativistic valley polarization, as illustrated in Fig. 1 (d).

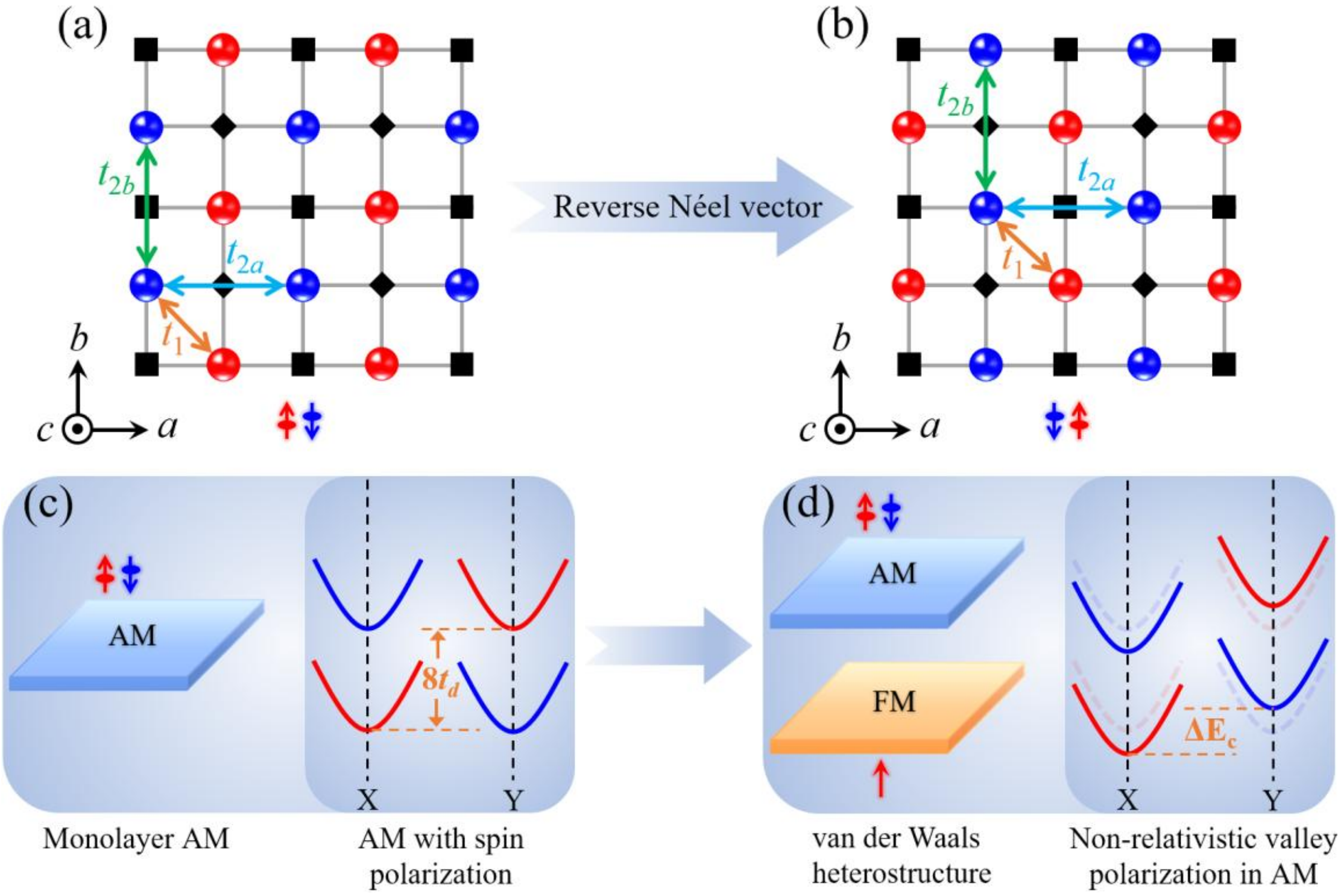

Fig. 1. (a) and (b) The schematic illustrations of the Lieb lattice of 2D tetragonal AM with opposite Néel vector, where red and blue balls represent magnetic atoms with opposite spins in different sublattices, squares and rhombus mark different non-magnetic atoms. (c) Schematic of the monolayer AM and the corresponding spin-polarized band structure. (d) vdW heterostructure formed by a monolayer AM and a FM, and the corresponding non-relativistic valley polarization ($\Delta E_c$) in the AM is marked by orange.

Benefiting from the flexibility and easy tunability of vdW heterostructures, both theory and experiments have demonstrated that vdW heterostructures composed of non-magnetic TMD and FM can exploit the magnetic proximity effect of FM to induce valley polarization in TMD[41–45]. Moreover, first-principles calculations indicate that the direction of magnetization of the FM couples with the valley polarization of the TMD, forming a strong magneto-valley coupling effect. It is worth noting that the formation of such valley polarization requires consideration of the SOC effect. Similarly, in vdW heterostructures comprising monolayer AM and FM, flipping the magnetization of the FM also flips the valley polarization at the X and Y points in the AM, as schematically shown in Fig. 2 (a) and 2 (b). Unlike heterostructures based on non-magnetic TMD, AM possess properties analogous to antiferromagnets, flipping of the Néel vector of the AM reverses the local magnetic moments of the magnetic atoms, yielding two distinct magnetic states. Combined with the two magnetic states arising from magnetization flipping in the FM, this gives rise to a total of four valley-polarized states (as illustrated in Fig. 2), suggesting that AM/FM vdW heterostructures hold potential application in the field of multistate storage.

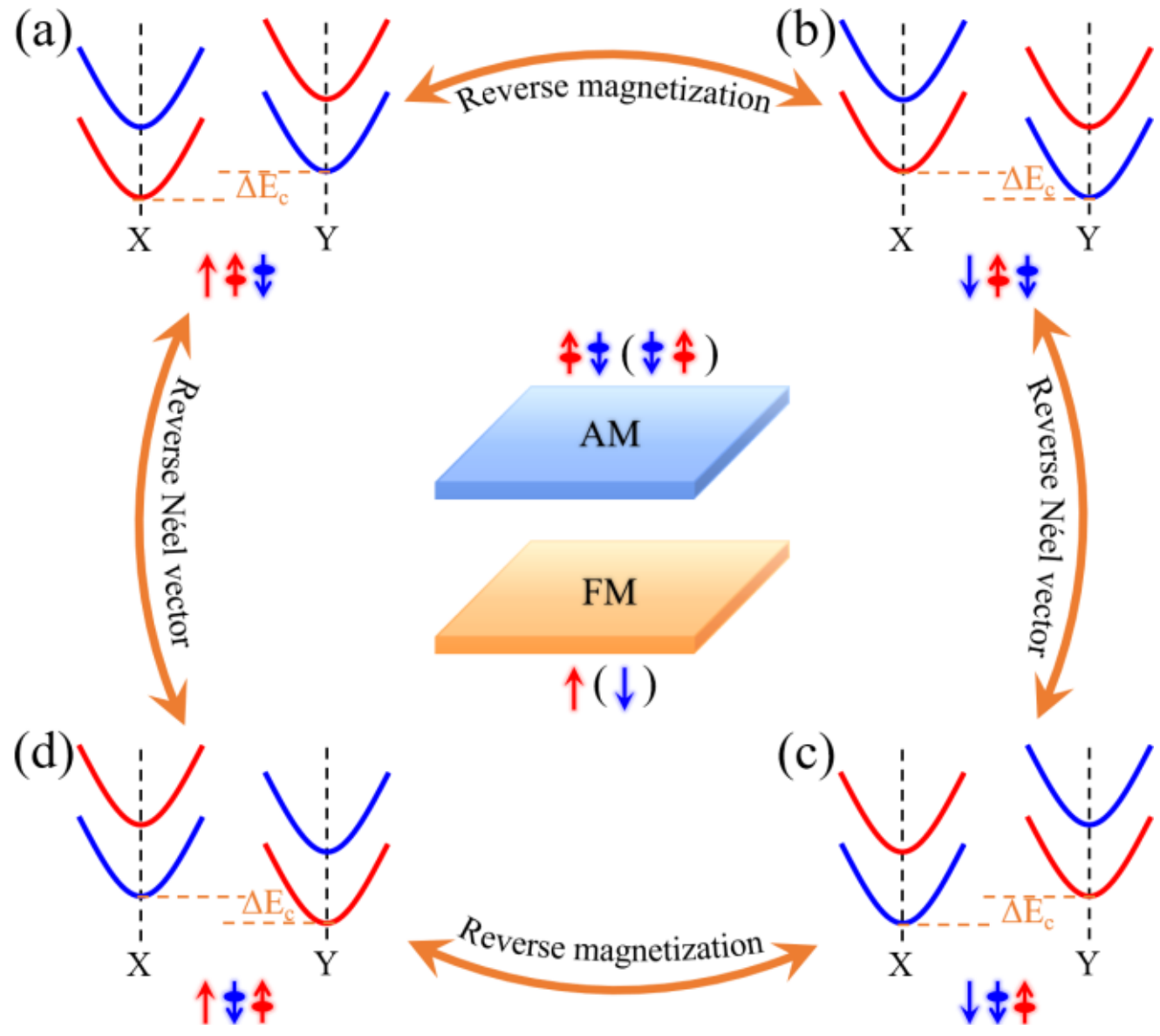

Fig. 2. Four independent valley-polarized states in the vdW heterostructure formed by a monolayer AM and a FM, arising from flipping the magnetization of the FM and the Néel vector of the AM.

We can still elucidate the four valley-polarized states shown in Fig. 2 from the above tight-binding model of proximity-induced valley polarization. From Eq. (6), it is clear that when the magnetization of the FM is along the +*z* direction, it is defined that $B_{eff} > 0$. If the X valley of the AM is contributed by spin-up electrons, then $s_X = +1$, the valley polarization of LCB is $\Delta E_c = -2B_{eff}$, corresponding to Fig. 2 (a). When the magnetization of the FM is reversed, $B_{eff} \rightarrow -B_{eff}$, the Néel vector of the AM is not flipped, the spin at the X valley remains $s_X = +1$, resulting in $\Delta E_c = 2B_{eff}$, corresponding to Fig. 2 (b). When both the magnetization of FM and the Néel vector of AM are reversed simultaneously, $B_{eff} \rightarrow -B_{eff}$ and $s_X = -1$, giving $\Delta E_c = -2B_{eff}$, corresponding to Fig. 2 (c). Note that reversing the Néel vector of the AM implies $t_d \rightarrow -t_d$ in Eq. (4), which interchanges the spins of electrons at the X and Y valleys, as illustrated in Fig. 2 (a) and (d), (b) and (c). If the magnetization of FM is along +*z* axis ($B_{eff} > 0$) while the X valley is contributed by spin-down electrons ($s_X = -1$), then $\Delta E_c = 2B_{eff}$, corresponding to Fig. 2 (d). This explains the reasonability of the four valley-polarized states illustrated in Fig. 2.

To date, numerous monolayer AM have been theoretically identified, most of which are semiconductors with a tetragonal lattice and exhibit a piezovalley effect under uniaxial strain[20,28,29]. However, tetragonal ferromagnetic semiconductors are relatively scarce. Studies have shown that monolayer CrP and CrAs with a tetragonal lattice are ferromagnetic semiconductors[34–36] and their lattice constants match well with those of most monolayer AM, making them highly suitable for constructing vdW heterostructures with AM. In this Letter, we take the monolayer FM CrP and the AM $V_2S_2O$ as representative examples. The two form a vdW heterostructure to investigate the non-relativistic valley polarization induced by the magnetic proximity effect. First-principles calculations reveal that CrP is a ferromagnetic semiconductor with a lattice constant of a = b = 4.17 Å and an indirect band gap of 41.04 meV, as shown in Fig. 3 (a) and (c). $V_2S_2O$ is an altermagnetic semiconductor with a lattice constant of a = b = 4.03 Å and an indirect band gap of 1.00 eV[46], as shown in Fig. 3 (b) and (d). The lattice mismatch between the two is approximately 3%, which falls within an acceptable range for constructing heterostructures.

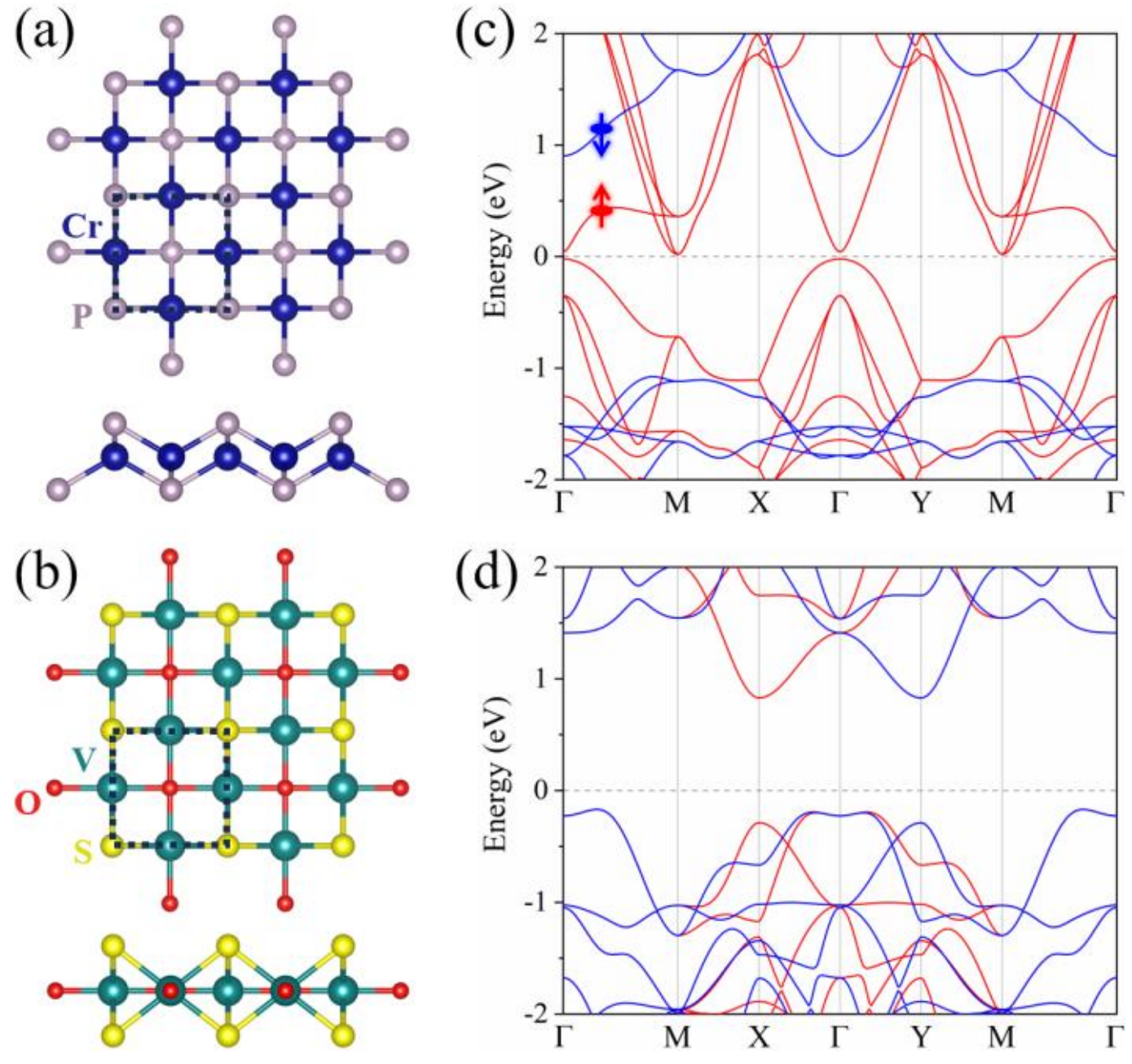

Fig. 3. Top and side views of the monolayers (a) CrP and (b) $V_2S_2O$, where the black dashed box denotes the unit cell. Spin-polarized band structures of (c) CrP and (d) $V_2S_2O$. The Fermi level is set to 0 eV in the band structures.

The vdW heterostructure composed of CrP and $V_2S_2O$ exhibits three distinct stacked configurations, as shown in Fig. S1 of the Supporting Information (SI). Calculations reveal that the stacked configuration in which the upper P atoms of CrP are located directly beneath the S atoms of $V_2S_2O$ possesses the lowest energy. The top and side views of the configuration are presented in Fig. 4 (a). Its lattice constants are a = b =

4.04 Å and the mirror symmetry along the diagonal direction is preserved. The corresponding band structure is shown in Fig. 4 (b). As can be seen from the band structure, the energy degeneracy of the LCB at the X and Y valleys near the Fermi level is lifted, giving rise to a non-relativistic valley polarization of approximately -53.03 meV. We calculated the spin density of the CrP and $V_2S_2O$ layers as well as the layer-resolved band structures after forming heterostructure, as shown in Fig. 4(c)-(f). Fig. 4 (c) and (e) indicate that the upper-layer $V_2S_2O$ retains rotational symmetry and exhibits altermagnetic characteristics, while the lower-layer CrP remains ferromagnetic. From Fig. 4 (d) and (f), it is clearly observed that the LCB at the X and Y valleys predominantly originate from contributions of the upper-layer AM $V_2S_2O$. Therefore, we conclude that it is precisely the presence of the magnetic proximity effect from the FM CrP that lifts the degeneracy at the X and Y valleys in the AM $V_2S_2O$, resulting in a remarkable non-relativistic valley polarization.

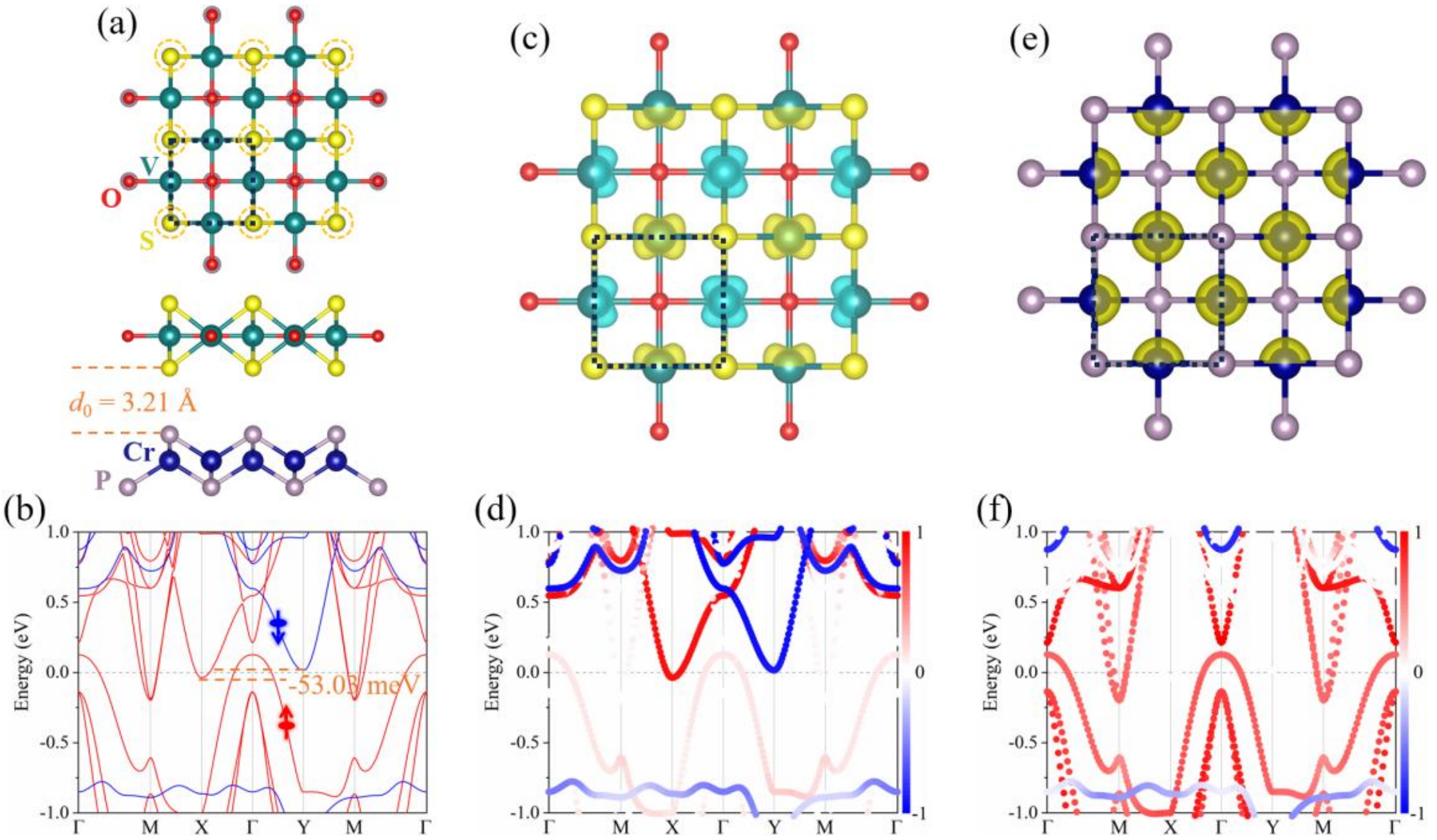

Fig. 4. (a) Top and side views of the most stable stacked configuration of the CrP/$V_2S_2O$ vdW heterostructure, where the black dashed box denotes the unit cell and the dashed circles indicate the upper P atoms in the CrP layer, and the interlayer distance of the heterostructure is marked in orange. (b) Band structure corresponding to the most stable configuration in (a). (c) Spin densities of the $V_2S_2O$ layer and (e) the CrP layer in the heterostructure, where the black dashed box denotes the unit cell and the isosurface value set to $5 \times 10^{-4}$ $e$/Bohr$^3$. Layer-resolved band contributions from the (d) $V_2S_2O$ layer and (f) CrP layer. The Fermi level is set to 0 eV in all band structures.

To demonstrate that the magnetic proximity effect is the fundamental origin of the non-relativistic valley polarization in the CrP/$V_2S_2O$ vdW heterostructure, we constructed vdW Waals heterostructures composed of the tetragonal non-magnetic monolayer PbO and AM $V_2S_2O$, as shown in Fig. S2 and S3 of the SI. The calculation results indicate that no valley polarization emerges in the LCB at the X and Y valleys contributed by $V_2S_2O$, whereas altermagnetic proximity effect appears only in the bands contributed by PbO layer. This phenomenon is consistent with the cases of PbO/$V_2Se_2O$ and SnO/$V_2Se_2O$ heterostructures[28,47,48]. Furthermore, we find that the valley polarization induced by the magnetic proximity effect is prevalent in such vdW heterostructures comprising monolayers AM and FM. (i) As shown in Fig. S1 of the SI, the metastable stacked configuration of the CrP/$V_2S_2O$ heterostructure also exhibits a non-relativistic valley polarization. (ii) We separately constructed lattice-matched CrAs/$V_2Se_2O$, CrAs/$Nb_2Se_2O$ and CrP/$Cr_2Te_2O$ vdW heterostructures and find that varying degrees of valley polarization occur in the most stable stacked configurations, as shown in Fig. S4-S6 of the SI. In addition, SOC has no significant impact on the valley polarization in the CrP/$V_2S_2O$ heterostructure (as shown in Fig. S7 of the SI). Therefore, it is concluded that the magnetic proximity effect from the FM is the underlying cause of the non-relativistic valley polarization in AM and this phenomenon is generalizable to FM/AM vdW heterostructures.

Tuning the interlayer distance via tension and compression is an effective means of modulating the properties of vdW heterostructures[43,49]. In Fig. S8 and S9 of the SI, we investigate the variation of valley polarization in the AM within the CrP/$V_2S_2O$ heterostructure with respect to the interlayer distance. As shown in Fig. S9 of the SI, stretching the interlayer distance overall reduces the magnitude of valley polarization. When stretched by 0.5 Å, the valley polarization decreases to 10.45 meV. In contrast, compressing the interlayer distance significantly enhances the valley polarization, which reaches 537.43 meV upon a compression of 0.5 Å. This also indirectly confirms that the magnetic proximity effect is the fundamental origin of the valley polarization in AM, and that the strength of the magnetic proximity effect (corresponding to $B_{eff}$ in Eq. (5)) impacts the magnitude of valley polarization in AM.

In heterostructures composed of FM and AM, the magnetization of the FM and the Néel vector of the AM are coupled to the valley polarization in the AM. Fig. 2 schematically illustrates that the heterostructure formed by a FM and an AM hosts four distinct valley-polarized states. Taking the CrP/$V_2S_2O$ heterostructure as an example, by reversing the magnetization of the FM CrP and the direction of Néel vector of the AM, we present the band structures of the four different states, as shown in Fig. 5 (e)-(h), with the corresponding schematic structures shown in Fig. 5 (a)-(d). As evident from Fig. 5(e)-(h), a strong coupling exists among the magnetization of CrP, the Néel vector of $V_2S_2O$ and the valley polarization, giving rise to four independent valley-polarized states. This materialization further solidate the validity of schematic illustration in Fig. 2 and demonstrates that heterostructures comprising FM and AM hold potential application in the field of valleytronics-based data storage.

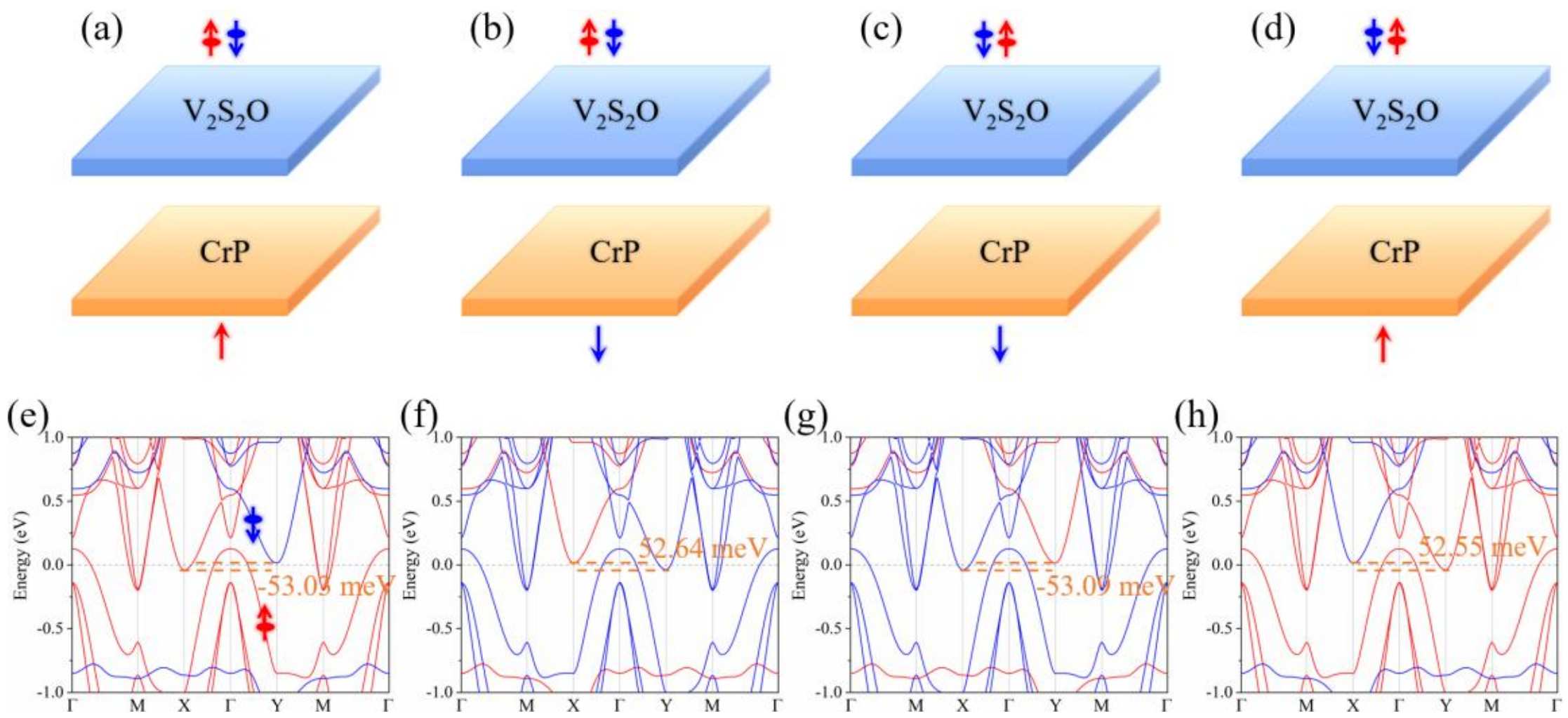

Fig. 5. (a)-(d) Schematic illustrations of the heterostructures formed by the FM CrP with different magnetizations and the AM $V_2S_2O$ with different Néel vector directions. (e)-(h) Four independent valley-polarized states in the CrP/$V_2S_2O$ vdW heterostructure, corresponding to (a)-(d), respectively, where the valley polarization of the LCB at the X and Y valleys are marked in orange. The Fermi level is set to 0 eV in all band structures.

## Conclusion

In summary, based on first-principles calculations, we have constructed various vdW heterostructures composed of monolayers FM and AM and find that the magnetic proximity effect from the FM induces pronounced non-relativistic valley polarization in the AM. Taking the CrP/$V_2S_2O$ heterostructure as an example, we reveal that the magnetic proximity effect of CrP induces a valley polarization of -53.03 meV in the bands contributed by $V_2S_2O$, and compressing the interlayer distance significantly enhances the magnitude of valley polarization. We further demonstrate that four independent valley-polarized states can be realized in the CrP/$V_2S_2O$ heterostructure by flipping the magnetization of CrP and the Néel vector of $V_2S_2O$. The non-relativistic valley polarization induced by the magnetic proximity effect and the phenomenon of four independent valley-polarized states are widely observed in FM/AM vdW heterostructures and qualitatively elucidated by a tight-binding model, suggesting that such heterostructures hold potential application in the field of valleytronics-based information storage.

## Acknowledgments

This work was supported by the National Natural Science Foundation of China (Grants No. 12504071), the Natural Science Foundation of Hunan Province (Grant No. 2026JJ60129), the Changsha Natural Science Foundation (Grant No. kg2502211), the Postdoctoral Fellowship Program of the China Postdoctoral Science Fund (Grant No. GZC20252247).